%% file: RDR.tex
\documentclass[sigconf, ,authorversion]{acmart}
\usepackage{enumitem}
\usepackage{bbm}
\usepackage{makecell}
\usepackage{multirow}
\usepackage{multicol}
\usepackage{booktabs}
\usepackage{url}
\usepackage{mathrsfs}
\usepackage{caption}
\usepackage{subfigure}
\usepackage{amsmath}

\usepackage{amssymb}
\usepackage{balance}

\copyrightyear{2024}
\acmYear{2024}
\setcopyright{acmlicensed}\acmConference[WWW '24 Companion]{Companion Proceedings of the ACM Web Conference 2024}{May 13--17, 2024}{Singapore, Singapore}
\acmBooktitle{Companion Proceedings of the ACM Web Conference 2024 (WWW '24 Companion), May 13--17, 2024, Singapore, Singapore}
\acmDOI{10.1145/3589335.3651551}
\acmISBN{979-8-4007-0172-6/24/05}
\begin{document}

\title{Recall-Augmented Ranking: Enhancing Click-Through Rate Prediction Accuracy with Cross-Stage Data }

\input{Abstract}



\author{Junjie Huang}
\orcid{0000-0002-5637-0735}
\affiliation{%
\institution{Shanghai Jiao Tong University}
\city{Shanghai}
\country{China}}
\email{huangjunjie2019@sjtu.edu.cn}

\author{Guohao Cai}
\orcid{0000-0002-9000-857X}
\affiliation{%
\institution{Huawei Noah's Ark Lab}
\city{Shenzhen}
\country{China}}
\email{caiguohao1@huawei.com}

\author{Jieming Zhu}
\orcid{0000-0002-5666-8320}
\affiliation{%
\institution{Huawei Noah's Ark Lab}
\city{Shenzhen}
\country{China}}
\email{jiemingzhu@ieee.org}

\author{Zhenhua Dong}
\orcid{0000-0002-2231-4663}
\affiliation{%
\institution{Huawei Noah's Ark Lab}
\city{Shenzhen}
\country{China}}
\email{dongzhenhua@huawei.com}

\author{Ruiming Tang}
\orcid{0000-0002-9224-2431}
\affiliation{%
\institution{Huawei Noah's Ark Lab}
\city{Shenzhen}
\country{China}}
\email{tangruiming@huawei.com}

\author{Weinan Zhang}
\orcid{0000-0002-0127-2425}
\affiliation{%
\institution{Shanghai Jiao Tong University}
\city{Shanghai}
\country{China}}
\email{wnzhang@sjtu.edu.cn}

\author{Yong Yu}
\orcid{0000-0003-0281-8271}
\affiliation{%
\institution{Shanghai Jiao Tong University}
\city{Shanghai}
\country{China}}
\email{yyu@sjtu.edu.cn}
\renewcommand{\shortauthors}{Junjie Huang et al.}

\maketitle

\input{Introduction}
\input{Background}
\input{Approach}

\input{Experiment}
\input{Conclusion}

\balance
\bibliographystyle{ACM-Reference-Format}
\bibliography{RDR}
\end{document}

%% file: Abstract.tex
\begin{abstract}

Click-through rate (CTR) prediction plays an indispensable role in online platforms. Numerous models have been proposed to capture users' shifting preferences by leveraging user behavior sequences.
However, these historical sequences often suffer from severe homogeneity and scarcity compared to the extensive item pool. Relying solely on such sequences for user representations is inherently restrictive, as user interests extend beyond the scope of items they have previously engaged with.
To address this challenge, we propose a data-driven approach to enrich user representations.
We recognize user profiling and recall items as two ideal data sources within the cross-stage framework, encompassing the u2u (user-to-user) and i2i (item-to-item) aspects respectively. 
In this paper, we propose a novel architecture named Recall-Augmented Ranking (RAR).
RAR consists of two key sub-modules, which synergistically gather information from a vast pool of look-alike users and recall items, resulting in enriched user representations.
Notably, RAR is orthogonal to many existing CTR models, allowing for consistent performance improvements in a plug-and-play manner. 
Extensive experiments are conducted, which verify the efficacy and compatibility of RAR against the SOTA methods.

\end{abstract}

%% file: Introduction.tex
\section{Introduction}
Recommender systems have been widely deployed to save users from information overload. Among them, CTR prediction is an essential task, which is to predict the probability that a user will click on an item under a particular context, enhancing both user experience and platform revenue. 

\begin{figure}[!tbp]
    \centering
     \includegraphics[scale=0.4]{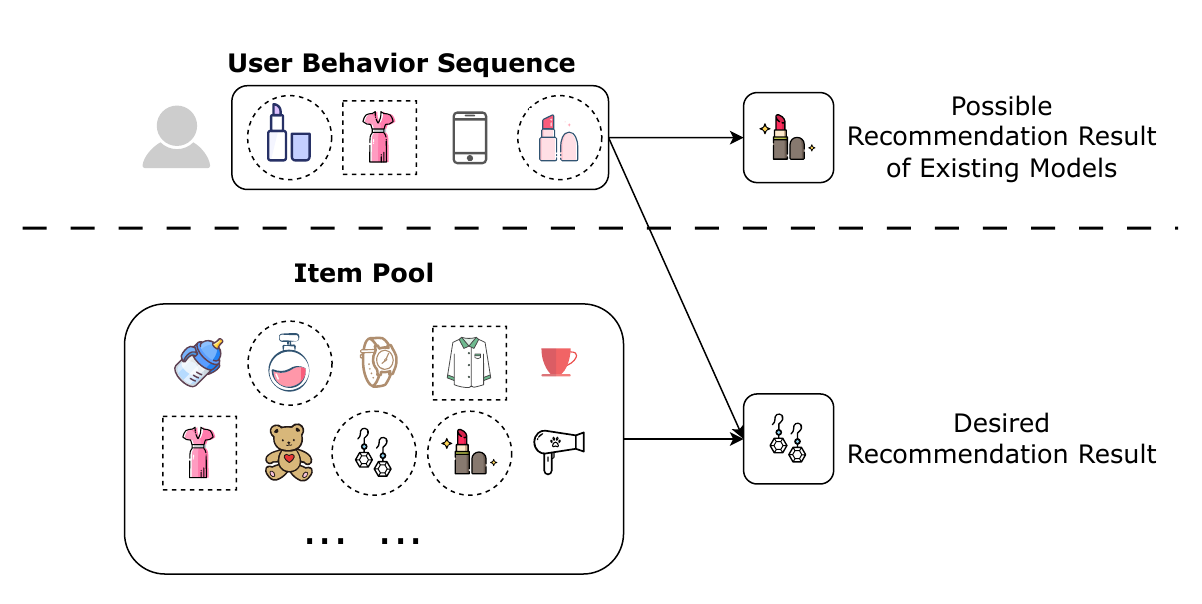}
    \caption{An illustrated example for motivations of RAR. 
    }
    \label{fig:intro}
\vspace{-1.6em}
\end{figure}

Recently, many models have been proposed to extract user interest based on historical behavior sequences. However, items in user behavior sequences often exhibit homogeneity and scarcity versus the large-scale item pool, which is detailed in Section 2.
Moreover, existing models often rely on target attention mechanisms, assigning higher scores to repetitive, similar items, reinforcing a cycle of homogeneity. 
In Figure~\ref{fig:intro}, we provide an example.
When a user buys lipstick, existing models often suggest similar products. However, the user might prefer exploring related items such as perfume or earrings, seeking variety beyond her initial purchase, even without previous interactions with these items.
Therefore, we aim to enrich user
representations from a data-driven perspective, incorporating diverse sources of information to enhance accuracy.

Moreover, CTR predictions traditionally focus on single user-item interactions and often overlook the interrelationships across various users and items, resulting in inadequate long-tail modeling.
On the contrary, the recall stage inherently generates similar user-item lists, providing cross-instance modeling capability.
We
recognize user profiling and recall items as two ideal data sources
within the cross-stage framework, encompassing the u2u and i2i aspects respectively.
 In this paper, we are interested in how to leverage these cross-stage data to enhance CTR prediction accuracy rather than how to construct the two sets.

In this paper, we propose a novel architecture named \textbf{Recall-Augemented Ranking (RAR)} to enhance model accuracy based on cross-stage data. RAR consists of two key components: the Cross-Stage User \& Item Selection Module and the Co-Interaction Module. These sub-modules efficiently gather information from a broad spectrum of look-alike users and recall items, thereby enriching user representations. Note that the Co-Interaction Module is a set-to-set modeling, which has not been previously explored in CTR prediction task. 

In summary, the contributions of the paper are as follows:
\begin{itemize}[topsep = 3pt,leftmargin =5pt]
\item We shed light on the limitations of relying solely on user behavior sequences to model user preferences. To address this inadequacy, we propose a novel architecture RAR, which leverages cross-stage data to enrich user representation.
\item RAR contains two extra data sources, namely the look-alike user set and recall item set. It is the first work that incorporates set-to-set modeling into CTR prediction to the best of our knowledge.
\item RAR serves as a framework capable of enhancing the performance of numerous existing CTR prediction models. Comprehensive experiments show RAR's outperformance, effectiveness and compatibility with a wide variety of models.
\end{itemize}

%% file: Background.tex
\section{BACKGROUND}
\begin{itemize}[leftmargin=*]
\item
\textbf{Motivation of RAR:}
We provide an analysis of user behavior in Taobao\footnote{\url{https://tianchi.aliyun.com/dataset/649}}.
Figure~\ref{fig:problem}(a) illustrates the scarcity of user historical sequences, with the majority of users having interacted with only a minuscule fraction of the total number of available items.
Figure~\ref{fig:problem}(b) highlights user behavior's homogeneity, with most activity of a specific user concentrated in four to five categories out of thousands.
Furthermore, traditional CTR models focus on single user-item interactions, yet overlook broader interrelationships. Conversely, the recall stage facilitates cross-instance modeling by linking similar user-item lists.
\begin{figure}[!tbp]
    \centering
     \includegraphics[scale=0.33]{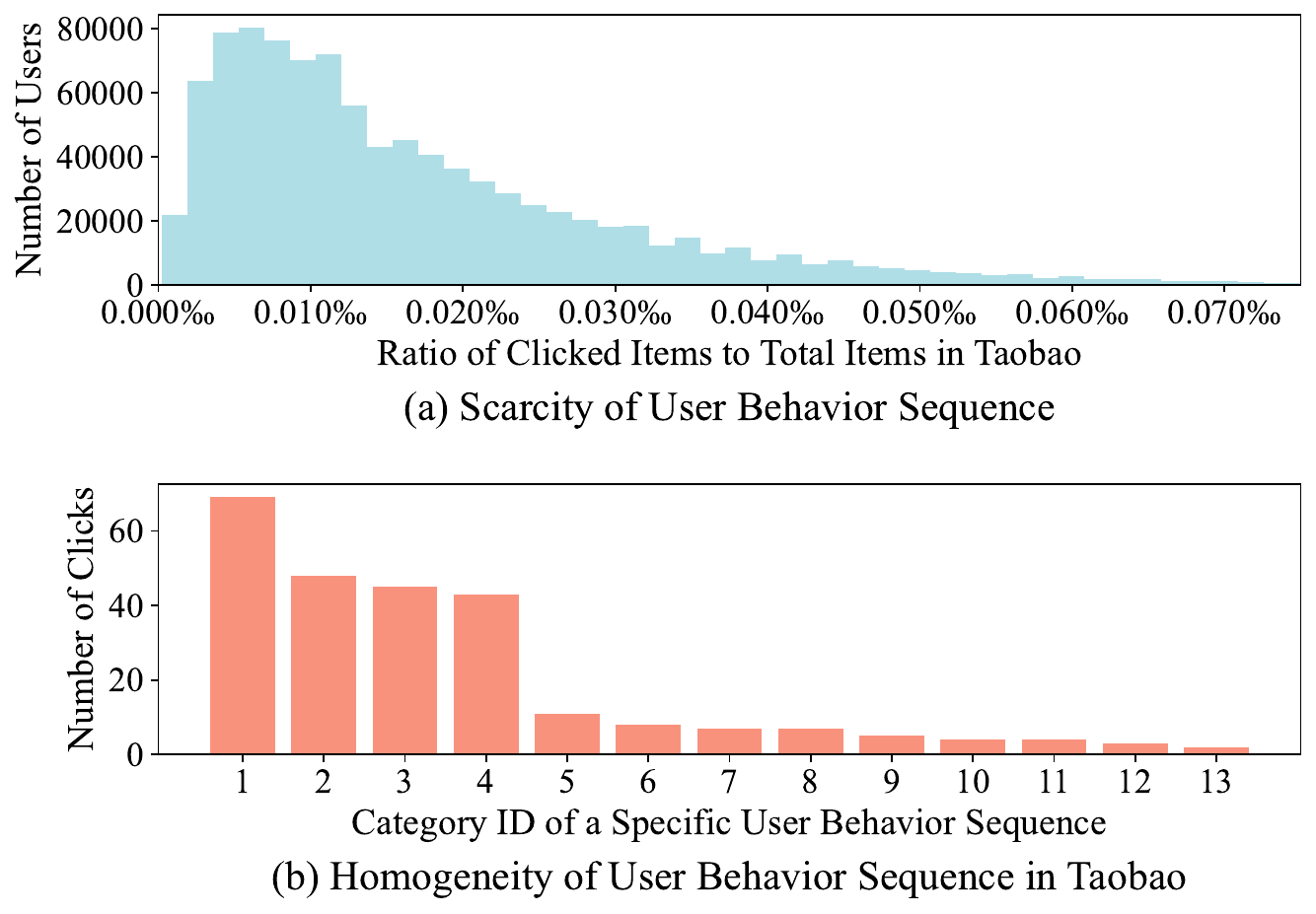}
    \caption{Observations of scarcity and homogeneity of user  behavior sequence in Taobao dataset.
    }
    \label{fig:problem}
\vspace{-1.1em}
\end{figure}
\item
\textbf{Cascade Ranking System and User Profiling:}
In modern information retrieval applications, a cascade ranking system is often used to balance the efficiency and effectiveness. The system includes a variety of rankers. Each stage selects the top-k items it receives and feeds them to the next stage. Among them, recall and ranking are two common stages. 
Besides, look-alike methods have become a core component of online advertising and marketing, which are intended to identify similar users from a small user set.
\end{itemize}

\input{tabels/notation}

%% file: tabels/notation.tex
\begin{table}[!t]
\renewcommand\arraystretch{1.2}
\setlength{\tabcolsep}{3.4pt}
\caption{Notations and descriptions}
\scalebox{0.66}
{
\begin{tabular}{c|l}
\hline Notation & Description. \\
\hline
$k_{l}, k_{r}$ & The number of selected 
look-alike users and recall items respectively. \\
$S_{\mathcal{L}}, S_{\mathcal{R}}$ & Similarity score matrix of look-alike users and recall items. \\
$E_{\mathcal{L}}, E_{\mathcal{R}}$ & Embedding matrix of look-alike users and recall items respectively. \\
$E_{\mathcal{L}^{'}}, E_{\mathcal{R}^{'}}$ & Embedding matrix of selected look-alike users and recall items. \\
$E_{\mathcal{L}H}^{'}, E_{\mathcal{R}H}^{'}$ & High-order representation of selected look-alike users and recall items. \\
$P$ & Hash projection matrix in selection modules. \\
\hline
\end{tabular}
}
\label{tab:notation}
\end{table}

%% file: Approach.tex
\section{Approach}

\subsection{Cross-Stage User/Item Selection Module}
The Cross-Stage User/Item Selection Module select the most similar users and relevant items. 
The selection process can be abstracted into two steps and we take the selection of recall items as an example. First, similarity is measured between the target item and each recall item by similarity function $f(\cdot)$.
Then top-k relevant recall items can be selected based on the similarity score, which can be formalized in Equation~\ref{equ:similarity}, ~\ref{equ:topk_retrieve}.
We conclude the key notations and the descriptions in Table~\ref{tab:notation}.
\begin{equation}\label{equ:similarity}
S_{\mathcal{L}}=f(E_{\mathcal{L}}, e^u),\quad
S_{\mathcal{R}}=f(E_{\mathcal{R}}, e^i)
\end{equation}
\begin{equation}\label{equ:topk_retrieve}
E_{\mathcal{L}^{'}}=[e^{u}_{k_{1}} e^{u}_{k_{2}} ... e^{u}_{k_{l}}]^T, \quad
E_{\mathcal{R}^{'}}=[e^{i}_{k_{1}} e^{i}_{k_{2}} ... e^{i}_{k_{r}}]^T
\end{equation}
An intuitive idea is using the embedding and search k-nearest neighbor by inner product. However, the huge number of multiplications makes real-world deployment impractical.  Considering the selection complexity, we use the SimHash function in our experiment.

SimHash, leveraging locality-sensitive properties, ensures similar outputs for similar inputs through random projection and signed axes, simplifying embeddings to binary fingerprints. This process is detailed in Equation~\ref{equ:simhash}, ~\ref{equ:simhash2}, where $\boldsymbol{e}^{i}_k$ stands for the embedding of the $k^{th}$ recall item and $m$ is the $m^{th}$ hash function in the hash function set. It reduces storage and speeds up selection by using hamming distance for efficient comparison. 

\begin{equation}\label{equ:simhash}
\operatorname{sig}^{i}_k[m]=\sum_{n=1}^{d_2} \operatorname{sgn}\left(\boldsymbol{e}^{i}_k[n] \cdot P[n][m]\right)
\end{equation}
\begin{equation}\label{equ:simhash2}
\operatorname{sig}^{i}_k[m]\xleftarrow{}\mathbbm{1}_{\operatorname{sig}^{i}_k[m]>0}(\operatorname{sig}^{i}_k[m])
\end{equation}
\begin{figure}[!thbp]
	\centering
	\includegraphics[width=0.47\textwidth]{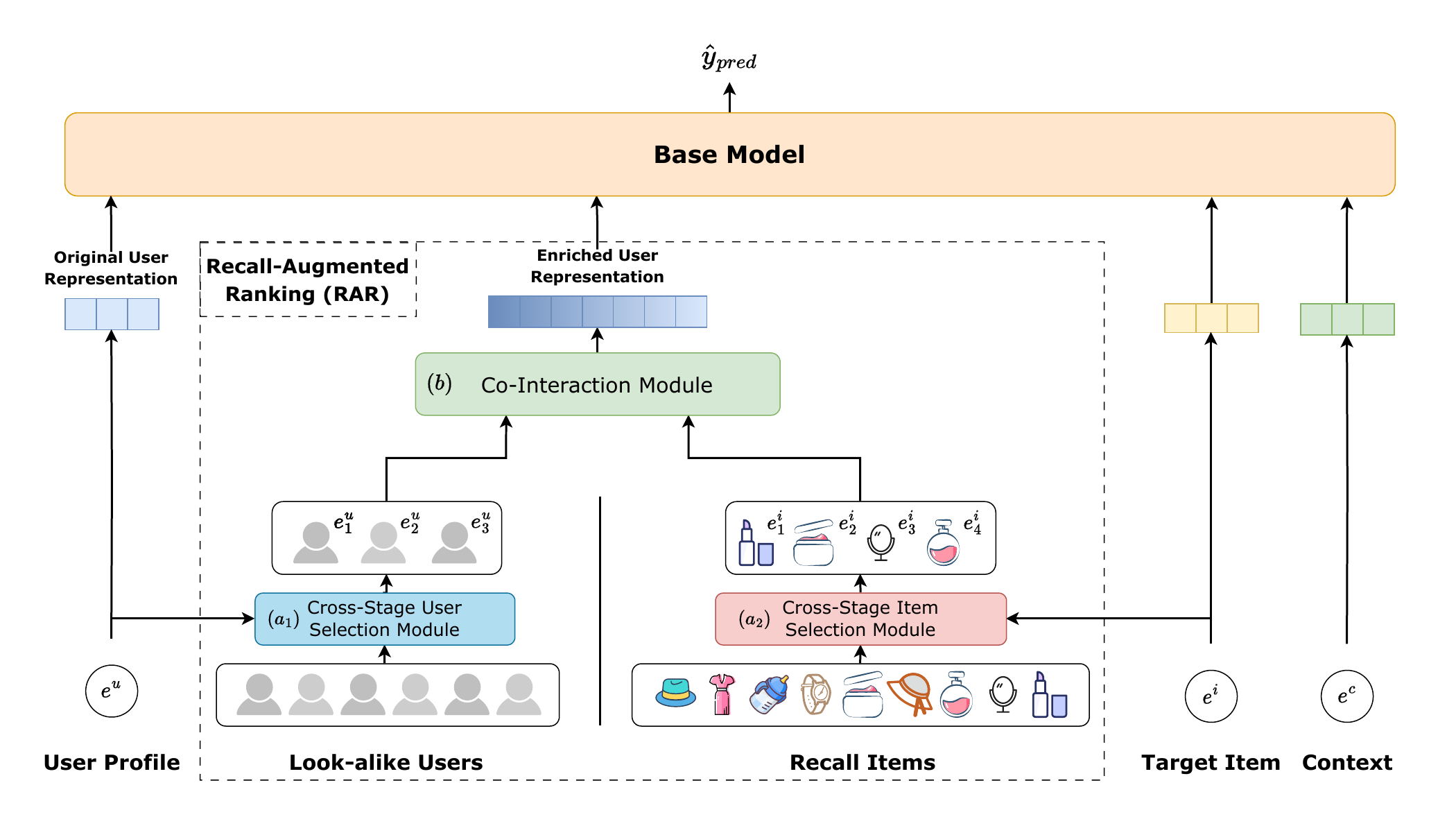}
	\caption{RAR applied in existing CTR prediction models. 
    }
	\label{fig:framework}
 \vspace{-0.6em}
\end{figure}
\subsection{Co-Interaction Module}
Co-Interaction Module provides a fine-grained set-to-set modeling. 
It improves upon the simplistic equal weighting of all selected recall items, which overlooks hierarchical information.
We introduce a matching matrix to assess user-item interest compatibility.
The matching score is represented as high-level latent vectors' inner product, as is shown in Equation~\ref{equ:high_order}.  
Then we compute the matching matrix in Equation~\ref{equ:matching_matrix}, where $Sigmoid(\cdot)$ is used to map the matching scores to (0,1). 
\begin{equation}\label{equ:high_order}
E_{\mathcal{L}H}^{'}=MLP(E_{\mathcal{L}}^{'}),\quad
E_{\mathcal{R}H}^{'}=MLP(E_{\mathcal{R}}^{'})
\end{equation}
\begin{equation}\label{equ:matching_matrix}
\mathcal{M}_{M}=Sigmoid(E_{\mathcal{L}H}^{'}\cdot E_{\mathcal{R}H}^{'T})
\end{equation}

To provide the model with a clearer indication of which recall items are more important, the signal $y_{ui}^{ep}$ is utilized to supervise the training of the matching matrix. As the exposed signal is very sparse and we define $y_{ui}^{ep}$ in Equation~\ref{equ:exposed_signal}. 
\begin{equation}\label{equ:exposed_signal_hat}
\hat{y}_{ui}^{ep}= \mathcal{M}_{M}
\end{equation}
\begin{equation}\label{equ:exposed_signal}
y_{ui}^{ep}= \begin{cases}0, & i \text { has never been exposed to $u^{\prime}$ for $\forall$ $u^{\prime}$ in $\mathcal{L}_{u}^{\prime}$} \\ 1, & \text { otherwise }\end{cases}
\end{equation}

Finally, the matching matrix is averaged by row and by column to obtain the item and user weighting vector. We obtain user common interest $v^{c}_{ui}$ and user diverse interest $v^{d}_{ui}$ by multiplying weighting vectors with corresponding embeddings.
Then user enriched representation $v^{enr}_{u}$ is obtained by concatenating $v^{c}_{ui}$ and $v^{d}_{ui}$.
\begin{equation}\label{equ:weighting_vector}
w_{i}= Mean(\mathcal{M}_{M}, axis=0), \quad
w_{u}= Mean(\mathcal{M}_{M}, axis=1)
\end{equation}
\begin{equation}\label{equ:enriched_feature}
v^{c}_{ui}=w_{u}\cdot E_{\mathcal{L}^{'}},\quad
v^{d}_{ui}=w_{i}\cdot E_{\mathcal{R}^{'}}
\end{equation}
\begin{equation}\label{equ:potential_interest}
v^{enr}_{u}=Concat(v^{c}_{ui}, v^{d}_{ui})
\end{equation}

\subsection{Objective function}
The loss function of RAR can be represented by Equation~\ref{equ:loss_fn}, where $\mathcal{L}_{clk}$ aims to predict CTR accurately and $\mathcal{L}_{ep}$ aims to provide a clearer indication to the model of which recall items are most important. $\alpha\in[0, 1]$ is a tunable parameter for balancing the two losses. Both two losses are cross-entropy loss and supervise the training process in a point-wise manner.
All modules of RAR are trained jointly by minimizing the joint loss function on the training dataset.
\begin{equation}\label{equ:loss_fn}
\mathcal{L}=\alpha\cdot\mathcal{L}_{clk} + (1-\alpha)\cdot\mathcal{L}_{ep}
\end{equation}

\subsection{Complexity Analysis}
We analyze the efficiency of the RAR in this section. Co-Interaction Module first gets the matching matrix and weighting scores by multiplying the embedding vectors of selected look-alike users  and recall items, followed by a weighted sum  using the weighting vectors. Therefore, three matrix multiplications are needed, and the time complexity is $O(B\cdot  k_{l}\cdot  k_{r} \cdot d)$. As for the selection module, the time complexity depends on the selection method utilized. Since conducting $XOR$ and counting the number of bits in 1 in SimHash can be accomplished in $O(1)$, the overall time complexity of RAR is $O(B\cdot (l+r) \cdot d + B\cdot  k_{l}\cdot  k_{r} \cdot d)$, where $d=Max(d_{1}, d_{2})$. Since $k_{l} << l$, $k_{r} << r$, it can be approximated as $O(B\cdot (l+r) \cdot d)$.

%% file: Experiment.tex
\section{Experiments}

\subsection{Experimental Setup}
\textbf{Datasets.} 
We conduct experiments on three public datasets. KKBox is a challenge dataset for music recommendation.
Movielens contains users' tagging records on movies.
CandiCTR-Pub is a
publicly available industrial dataset that is both practical and large-scale. 
Apart from CandiCTR-Pub, which already includes recall item sets, we manually construct recall item sets and look-alike user sets for KKBox and MovieLens by employing a pretrained matching model(e.g., DSSM) for inner product calculations between user-item and user-user pairs. 
Our implementation builds upon FuxiCTR and follow the public benchmark~\cite{BARS} and previous works~\cite{FRNet, CIM, lin2023map}.

\textbf{Base models.} 
We consider both high-order feature interaction and ensemble models.
We choose IPNN, WDL, DeepFM, DCN, xDeepFM, AutoInt+, DeepIM, DCN-V2 as our base models, which has been evaluated in the BARS benchmark; see references in~\cite{BARS, zhu2021open}.

\textbf{Baselines.}  FRNet~\cite{FRNet} learns context-aware feature representations by capturing cross-feature relationships, becoming the new SOTA. CIM~\cite{CIM} encodes all candidate items into a context vector by transformer to characterize users' implicit awareness. 

\textbf{Metrics.} We apply the most popular metrics AUC and gAUC (weighted sum AUC, grouped by users) to evaluate the performance. 

\subsection{Performance Evaluation with SOTA Models}

\input{tabels/main_result}
We evaluate RAR on existing models, including many SOTA methods, which is shown in Table~\ref{tab:main_table}. 
RAR notably surpasses other methods, with xDeepFM+RAR improving AUC by up to 4.7\% across datasets, demonstrating the efficacy of using cross-stage data for richer user representations. 

\subsection{Ablation Study}
We investigate the effectiveness of different components of RAR in Table~\ref{tab:ablation}. Three typical base models are selected to ensure generalizability and fairness. 
\begin{itemize}[leftmargin=*]
\item 
\textbf{Removing the channel of look-alike users:}
\textbf{RAR-user} replaces the channel of look-alike users with the target user only. 
Table~\ref{tab:ablation} shows a 2.6\% gAUC and 2.5\% AUC increase over raw CTR models, highlighting the benefit of incorporating recall items into ranking models for diversified user representations. However, RAR-user's comparison to RAR indicates further potential by leveraging user common interest introduced by look-alike users.
\item 
\textbf{Removing the User and Item Selection Module:}
\textbf{RAR-select}, removing the Cross-Stage Selection Module, truncates the look-alike user set and recall item set to match RAR's scale. Table~\ref{tab:ablation} reveals a 0.4\% gAUC and 2.3\% AUC boost over base CTR models, which indicates the importance of a careful selection for further accuracy improvement.
\item 
\textbf{Removing the Co-Interaction Module:}
\textbf{RAR-aux-wght}, omitting the Co-Interaction Module and related losses, uses simple sumpooling for user interest representation and shows weaker performance. \textbf{RAR-wght}, dropping weighting vectors but keeping auxiliary loss, significantly outperforms RAR-aux-wght, highlighting the auxiliary loss's role in guiding hierachical information of recall items. RAR's superiority over RAR-wght underscores the value of both matching loss and weighting vectors.
\end{itemize}

\subsection{Training Efficiency}
We present a wall-time comparison of RAR and CIM, the latter of which has proven effective in a real-world search advertising system. 
Findings detailed in Table~\ref{tab:efficiency} reveal that CIM\_short, utilizing a truncated context input of 50 recall items, exhibits the least time consumption, whereas CIM\_long, processing the full set of 305 recall items, incurs the most time. 
 RAR's selection modules effectively filter noise from extensive recall pools and look-alike user sets without reducing information. 
 Additionally, as Section 4.2 demonstrates, the performance gain of RAR is substantial.

\input{tabels/table_ablation}

\input{tabels/efficiency}


%% file: tabels/main_result.tex
\begin{table}[!htb]
\renewcommand\arraystretch{1.1}
\setlength{\tabcolsep}{2.5pt}
\caption{Overall performance comparison against the state-of-the-art models on three datasets.}
\resizebox{0.47\textwidth}{!}{%
\begin{tabular}{ccccccccc}
\toprule
\multicolumn{1}{c}{Datasets} & \multicolumn{8}{c}{KKBox} \\
\cmidrule(lr){1-1}\cmidrule(lr){2-9}
\multicolumn{1}{c}{Modules} & \multicolumn{2}{c}{Raw} & \multicolumn{2}{c}{+FRNet} & \multicolumn{2}{c}{+CIM} & \multicolumn{2}{c}{+RAR} \\
\cmidrule(lr){1-1}\cmidrule(lr){2-3}\cmidrule(lr){4-5}\cmidrule(lr){6-7}\cmidrule(lr){8-9}
Models & gAUC(\%) & AUC(\%) & gAUC(\%) & AUC(\%) & gAUC(\%) & AUC(\%) & gAUC(\%) & AUC(\%) \\ \midrule
IPNN & 78.75 & 85.25 & 78.27  & 84.94 & 78.31 & 85.25 & 80.15 & 86.45\\
WDL & 78.44 & 85.02 &  78.26 & 84.85 & 78.67 & 85.36 & 79.74 & 86.23\\
DeepFM & 78.76 & \underline{85.32} & \underline{78.70}  & \underline{85.26} & 78.90 & 85.68 & 80.14 & 86.51 \\
DCN & 78.66 & 85.25 & 78.69 & \underline{85.26} & \underline{79.16} & \underline{85.74} & \underline{80.22} & \underline{86.58}  \\
xDeepFM & 78.60 & 85.25 &  78.65 & 85.22 & 78.72 & 85.56 & 80.12 & 86.50 \\
AutoInt+ & \underline{78.78} & \textbf{85.34} & \textbf{78.71}  & \textbf{85.28} & 78.94 & 85.64 & 80.20 & 86.55 \\
DeepIM & \textbf{78.79} & 85.29 & 78.56  & 85.16 & 78.92 & 85.63 & \textbf{80.26} & \textbf{86.59} \\
DCN-V2 & 78.64 & 85.17 & 78.62  & 85.22 & \textbf{79.45} & \textbf{85.77} &  80.12 & 86.49 \\
\cmidrule(lr){1-1}\cmidrule(lr){2-3}\cmidrule(lr){4-5}\cmidrule(lr){6-7}\cmidrule(lr){8-9}
Best RelImp & \multicolumn{1}{c}{0.0\%} & \multicolumn{1}{c}{0.0\%} & \multicolumn{1}{c}{0.1\%}  & \multicolumn{1}{c}{0.1\%} & \multicolumn{1}{c}{1.0\%} & \multicolumn{1}{c}{0.7\%} & \multicolumn{1}{c}{2.0\%} & \multicolumn{1}{c}{1.6\%}\\
\hline
\multicolumn{1}{c}{Datasets} & \multicolumn{8}{c}{Movielens}\\
\cmidrule(lr){1-1}\cmidrule(lr){2-9}
\multicolumn{1}{c}{Modules} & \multicolumn{2}{c}{Raw} & \multicolumn{2}{c}{+FRNet} & \multicolumn{2}{c}{+CIM} & \multicolumn{2}{c}{+RAR} \\
\cmidrule(lr){1-1}\cmidrule(lr){2-3}\cmidrule(lr){4-5}\cmidrule(lr){6-7}\cmidrule(lr){8-9}
Models & gAUC(\%) & AUC(\%) & gAUC(\%) & AUC(\%) & gAUC(\%) & AUC(\%) & gAUC(\%) & AUC(\%) \\ \midrule
IPNN & \textbf{95.53} & \textbf{96.53} &  95.14 & 96.15 & 95.28  & 96.38 &  \textbf{95.92} & \textbf{97.02} \\
WDL & 95.29 & 96.23 & 95.17 & 96.19 & \textbf{95.36} & \textbf{96.44} & 95.73  & 96.67 \\
DeepFM & 94.84 & 95.90 & 94.65 & 96.11  &  95.06  & 96.23 & 95.40 & 96.40  \\
DCN & \underline{95.32} & \underline{96.35} & 95.21 & \textbf{96.33} & 95.30 & 96.37 &  95.51 & 96.54 \\
xDeepFM & 95.27 & 96.20 & 95.21 & 96.26 &  95.16 & 96.29 & \underline{95.82} & \underline{96.81} \\
AutoInt+ & 95.22 & 96.24 & \textbf{95.26} & \underline{96.28} & 95.31 & 96.38 & 95.61 & 96.58\\
DeepIM & 95.29 & 96.29 & 95.21 & \underline{96.28} & 95.30 & \underline{96.39} & 95.51 & 96.61 \\
DCN-V2 & 94.95 & 96.00 & \underline{95.23} & 96.25 & \underline{95.35} & \underline{96.39} & 95.66 & 96.63\\
\cmidrule(lr){1-1}\cmidrule(lr){2-3}\cmidrule(lr){4-5}\cmidrule(lr){6-7}\cmidrule(lr){8-9}
Best RelImp & \multicolumn{1}{c}{0.0\%} & \multicolumn{1}{c}{0.0\%} & \multicolumn{1}{c}{0.3\%}  & \multicolumn{1}{c}{0.3\%} & \multicolumn{1}{c}{0.4\%} & \multicolumn{1}{c}{0.4\%} & \multicolumn{1}{c}{0.7\%} & \multicolumn{1}{c}{0.7\%}\\
\hline
\multicolumn{1}{c}{Datasets} & \multicolumn{8}{c}{CandiCTR-Pub}\\
\cmidrule(lr){1-1}\cmidrule(lr){2-9}
\multicolumn{1}{c}{Modules} & \multicolumn{2}{c}{Raw} & \multicolumn{2}{c}{+FRNet} & \multicolumn{2}{c}{+CIM} & \multicolumn{2}{c}{+RAR} \\
\cmidrule(lr){1-1}\cmidrule(lr){2-3}\cmidrule(lr){4-5}\cmidrule(lr){6-7}\cmidrule(lr){8-9}
Models & gAUC(\%) & AUC(\%) & gAUC(\%) & AUC(\%) & gAUC(\%) & AUC(\%) & gAUC(\%) & AUC(\%) \\ \midrule
IPNN & \textbf{52.87} & 60.92 & 52.35 & \textbf{61.08} & 53.76 & 61.86 & 54.35 & 62.53 \\
WDL & \underline{52.82} & 60.90 & 52.48 & 60.82 & 53.92 & 62.73 & 54.43 & \underline{63.92}\\
DeepFM & \underline{52.82} & 60.99 & 52.48 & 60.89 & \textbf{53.94} & \underline{62.80} & \textbf{54.50} & \underline{63.92} \\
DCN & 52.78 & 60.95 & 52.55 & 60.61 & 53.75 & 61.75 & \textbf{54.50} & 62.53 \\
xDeepFM & \textbf{52.87} & \underline{61.19} & 52.48 & 60.78 & \underline{53.93} & 62.79 & 54.40 & \textbf{64.06} \\
AutoInt+ & 52.61 & 61.10 & \textbf{52.73} & \underline{60.95} & 53.82 & \textbf{62.85} & \underline{54.47} & 63.86\\
DeepIM & 52.72 & \textbf{61.23} & 52.42 & 60.94 & 53.48 & 61.65 & 54.44 & 62.72\\
DCN-V2 & 52.65 & 61.06 & \underline{52.64} & 60.79 & 53.32 & 61.69 & 54.30 & 62.70 \\
\cmidrule(lr){1-1}\cmidrule(lr){2-3}\cmidrule(lr){4-5}\cmidrule(lr){6-7}\cmidrule(lr){8-9}
Best RelImp & \multicolumn{1}{c}{0.0\%} & \multicolumn{1}{c}{0.0\%} & \multicolumn{1}{c}{0.2\%}  & \multicolumn{1}{c}{0.3\%} & \multicolumn{1}{c}{2.6\%} & \multicolumn{1}{c}{3.0\%} & \multicolumn{1}{c}{3.5\%} & \multicolumn{1}{c}{5.0\%}\\
\bottomrule
\end{tabular}
}
\vspace{-3pt}
\label{tab:main_table}
\end{table}

%% file: tabels/table_ablation.tex
\begin{table}[!t]
\vspace{-0.6em}
\renewcommand\arraystretch{1}
\setlength{\tabcolsep}{3.4pt}
\caption{Ablation study of RAR on CandiCTR-Pub.}
\scalebox{0.65}
{
\begin{tabular}{ccccccccc}
\toprule
\multirow{2}{*}{Model} & \multicolumn{2}{c}{DCN-V2} & \multicolumn{2}{c}{xDeepFM} & \multicolumn{2}{c}{DeepIM} & \multicolumn{2}{c}{Average RelImp}\\ \cmidrule(lr){2-3}\cmidrule(lr){4-5}\cmidrule(lr){6-7}\cmidrule(lr){8-9}
 & gAUC(\%) & AUC(\%) & gAUC(\%) & AUC(\%) & gAUC(\%) & AUC(\%) & gAUC(\%) & AUC(\%)\\ \hline
 Raw & 52.65 & 61.06 & 52.87 & 61.19 & 52.72 & 61.23 & 0\% & 0\%\\
 RAR-user & 54.07 & 62.35 & 54.09 & 63.45 & 54.19 & 62.23 & 2.6\% & 2.5\%\\
 RAR-select & 52.97 & 62.11 & 52.99 & 63.53 & 52.96 & 62.03 & 0.4\% & 2.3\%\\
 RAR-aux-wght & 52.48 & 61.14 & 52.57 & 62.52 & 52.49 & 61.10 & -0.4\% & 0.7\%\\
 RAR-wght & 53.89 & 62.47 & 53.87 & 63.78 & 54.06 & 62.35 & 2.3\% & 2.8\%\\
 \hline
 \textbf{RAR} & \textbf{54.30} & \textbf{62.70} & \textbf{54.40} & \textbf{64.06} & \textbf{54.44} & \textbf{62.72} & \textbf{3.1\%} & \textbf{3.3\%}\\
 \bottomrule
\end{tabular}
}
\label{tab:ablation}
\end{table}

%% file: tabels/efficiency.tex
\begin{table}[!t]
\vspace{-0.6em}
\renewcommand\arraystretch{1.2}
\setlength{\tabcolsep}{3.4pt}
\caption{Wall time comparison of the training and inference time of RAR and CIM on CandiCTR-Pub. }
\resizebox{0.9\linewidth}{!}
{
\begin{tabular}{ccccc}
\toprule
Model & Training Time & Rel.Inc & Time per inference step & Rel.Inc\\ 
\hline
CIM\_short & $\sim$14.5 min & 0\% & $\sim$18 ms & 0\%\\
CIM\_long & $\sim$29 min & 100\% & $\sim$48 ms & 167\%\\
RAR & $\sim$18 min & 24\% & $\sim$27.5 ms & 53\%\\
 \bottomrule
\end{tabular}
}
\label{tab:efficiency}
\vspace{0.6ex}
\end{table}

%% file: Conclusion.tex
\section{CONCLUSION}
In this paper, we first
shed light on the limitations of relying solely on homogeneous user behavior
sequences to model user preferences and then we propose a novel architecture called RAR which utilizes cross-stage data to improve the cross-instance modeling capability of the models.
RAR consists of two key sub-modules, which synergistically gather information from a vast pool of look-alike users and recall items, resulting in enriched user representations. RAR is a general framework that demonstrates great performance and compatibility through our in-depth experiments.

\begin{acks}
The Shanghai Jiao Tong University team is partially supported by National Natural Science Foundation of China (62177033). We also gratefully acknowledge the support of MindSpore\footnote{\url{https://www.mindspore.cn/}}, which is a
new deep learning computing framework used for this research.
\end{acks}